\documentclass[aps, prx,twocolumn,longbibliography,superscriptaddress,showpacs,amsmath,amssymb]{revtex4}
\usepackage[latin9]{inputenc}
\usepackage{amssymb}
\usepackage{graphicx}
\usepackage{amsmath}
\usepackage{color}
\usepackage{mathrsfs}
\usepackage{float}
\usepackage{indentfirst}
\usepackage{mathrsfs}
\usepackage{float}
\usepackage{indentfirst}
\usepackage{textcomp}
\usepackage{comment}
\usepackage{mathtools}
\usepackage{natbib,hyperref}
\usepackage{soul}

\begin{document}

\title{Self-consistent optimization of the trial wave-function in constrained path auxiliary field Quantum Monte Carlo using mixed estimators}

\author{Mingpu Qin} \thanks{qinmingpu@sjtu.edu.cn}
\affiliation{Key Laboratory of Artificial Structures and Quantum Control (Ministry of Education),  School of Physics and Astronomy, Shanghai Jiao Tong University, Shanghai 200240, China}

\date{\today}

\begin{abstract}
We propose a new scheme to implement the self-consistent optimization of the trial wave-function in constrained path auxiliary field Quantum Monte Carlo (CP-AFQMC) in the framewok of natural orbitals. In this scheme, a new trial
wave-function in the form of Slater determinant is constructed from the CP-AFQMC results by diagonalizing the mixed estimator of the one-body reduced density matrix. We compare two ways (from real and mixed estimators in CP-AFQMC) to calculate the one-body reduced density matrix in the self-consistent process and study the ground state of doped two dimensional Hubbard model to test the accuracy of the two schemes. By comparing the local density, occupancy, and ground state energy we find the scheme in which one-body reduced density matrix is calculated from mixed estimator is computational more efficient and provides more accurate result with less fluctuation. The local densities from mixed estimator scheme agree well with the numerically exact values. This scheme provides a useful tool for the study of strongly correlated electron systems.     
\end{abstract}

%\pacs{71.10.Fd, 02.70.Ss, 05.30.Fk}

\maketitle

\section{introduction}
Exotic quantum states can emerge from strongly correlated quantum many-body systems which largely enrich our understanding of quantum phases and the transitions between them \cite{RevModPhys.66.763}. But the studies of these systems mainly rely on a variety of numerical methods nowadays because analytic solution is rare \cite{PhysRevLett.20.1445}. Different numerical methods \cite{PhysRevX.5.041041} were developed in the past few decades for different types of systems. Quantum Monte Carlo (QMC) \cite{PhysRevD.24.2278,SUGIYAMA19861,PhysRevB.40.506,RevModPhys.83.349} is a very efficient method when the negative sign problem \cite{PhysRevB.41.9301,PhysRevLett.94.170201} is absent. However, most realistic quantum problems suffer from the negative sign problem which prevents the reach of low temperature and large system sizes. Constrained path Auxiliary Field Quantum Monte Carlo (CP-AFQMC) \cite{Zhang97} was developed to attack the negative sign problem in the study of Fermion systems with QMC. The essence of CP-AFQMC is the introduction of a trial wave-function, with which the sampling process in QMC is modified to avoid the negative sign problem. The price to pay is the introduction of systematic error (constraint error) in the final results. The quality of trial wave-function determines the accuracy of CP-AFQMC results.  Trial wave-functions were usually chosen empirically \cite{PhysRevB.78.165101,PhysRevB.88.125132,PhysRevB.89.125129,PhysRevB.94.085103} and previous benchmarks on different systems show the constraint errors are modest in many cases.

 In 2016, a self-consistent approach for the optimization of the trial wave-function was developed \cite{PhysRevB.94.235119}. In this approach, the QMC calculation is coupled with a mean-field Hamiltonian to get rid of the dependence of QMC results on the choice of the initial trial wave-function and to reduce the constraint error \cite{PhysRevB.94.235119}. Employing this self-consistent CP-AFQMC approach together with other state-of-art methods, the ground state of underdoped Hubbard model was determined to be a stripe phase \cite{Zheng1155}. This self-consistent approach was also generalized to the finite temperature calculations \cite{PhysRevB.99.045108}. After that, another self-consistent scheme in the framework of pseudo-BCS wave-function was also developed \cite{PhysRevResearch.3.013065}. The self-consistent strategy was also implemented in the real material calculations \cite{doi:10.1063/5.0031024}. Recently, the phase diagram of the two dimensional doped Hubbard model regarding stripe and spin density wave was determined with the self-consistent CP-AFQMC approach \cite{PhysRevResearch.4.013239}. 

In \cite{PhysRevB.94.235119}, besides the scheme to couple the CP-AFQMC calculation to a mean-filed Hamiltonian, it was also proposed to construct new trial wave-function with natural orbitals from CP-AFQMC in the self-consistent process. However, it was found that the fluctuation in this scheme using natural orbitals is very large, especially for the charge density (results not published in \cite{PhysRevB.94.235119}, can be found in the results section below).  
In this work, we propose a new scheme to implement the self-consistent optimization of the trial wave-function in the framework of natural orbitals in CP-AFQMC by taking advantage of mixed estimators. In this framework, we start by choosing an initial trial wave-function and perform one step of CP-AFQMC calculation using the initial wave-function. We then build a new trial wave-function in the form of Slater determinant from the natural orbitals (eigenvectors of the one-body reduced density matrix) of CP-AFQMC results. This new trial wave-function is then utilized in the next step CP-AFQMC calculation and the process is repeated until physical quantities are converged. 
%The idea here is similar as in \cite{PhysRevB.94.235119} with the difference lying in the way to construct new trial wave-function from QMC results. In \cite{PhysRevB.94.235119}, new trial wave-function is obtained by optimizing a mean-filed Hamiltonian.
We compare two schemes to calculate the one-body reduced density matrix (1-RDM), i.e., with mixed and real estimators. In usual CP-AFQMC calculation, mixed estimator is used to calculate physical quantities commuting with Hamiltonian (energy, for example). For other quantities, the mixed estimators are biased and we need to implement the so called back-propagation \cite{Zhang97,PhysRevE.70.056702} to calculate the real estimators for them. The calculation of mixed estimator is computational more efficient and with less fluctuation than the real estimator. However, from the benchmark calculations of the doped two dimensional Hubbard model with the two schemes, we find the one with mixed estimator is more accurate and more efficient. By analyzing the occupancy of natural orbitals in the two schemes, we find that all the information from mixed estimator of 1-RDM is contained in the new trial wave-function (see the appendix for a rigorous proof), while truncation is needed when 1-RDM is calculated with real estimator. This explains why the self-consistent scheme with mixed estimator is more accurate. Comparing to the previous self-consistent schemes \cite{PhysRevResearch.3.013065}, the new scheme is computational much cheaper because only mixed estimator is needed. Another advantage of the new scheme is that there is no tunable parameter in the calculation which makes the results more reliable.

The rest of the paper is organized as follows: In Sec.~\ref{model_method},
we give a brief review of the CP-AFQMC method. We then show the
comparison of the results with two schemes in Sec.~\ref{result}. We compare the local spin and hole densities,
the occupancy of the natural orbitals and the ground energy.
We summarize this work in Sec.~\ref{sum}.

\begin{figure*}[t]
	\includegraphics[width=80mm]{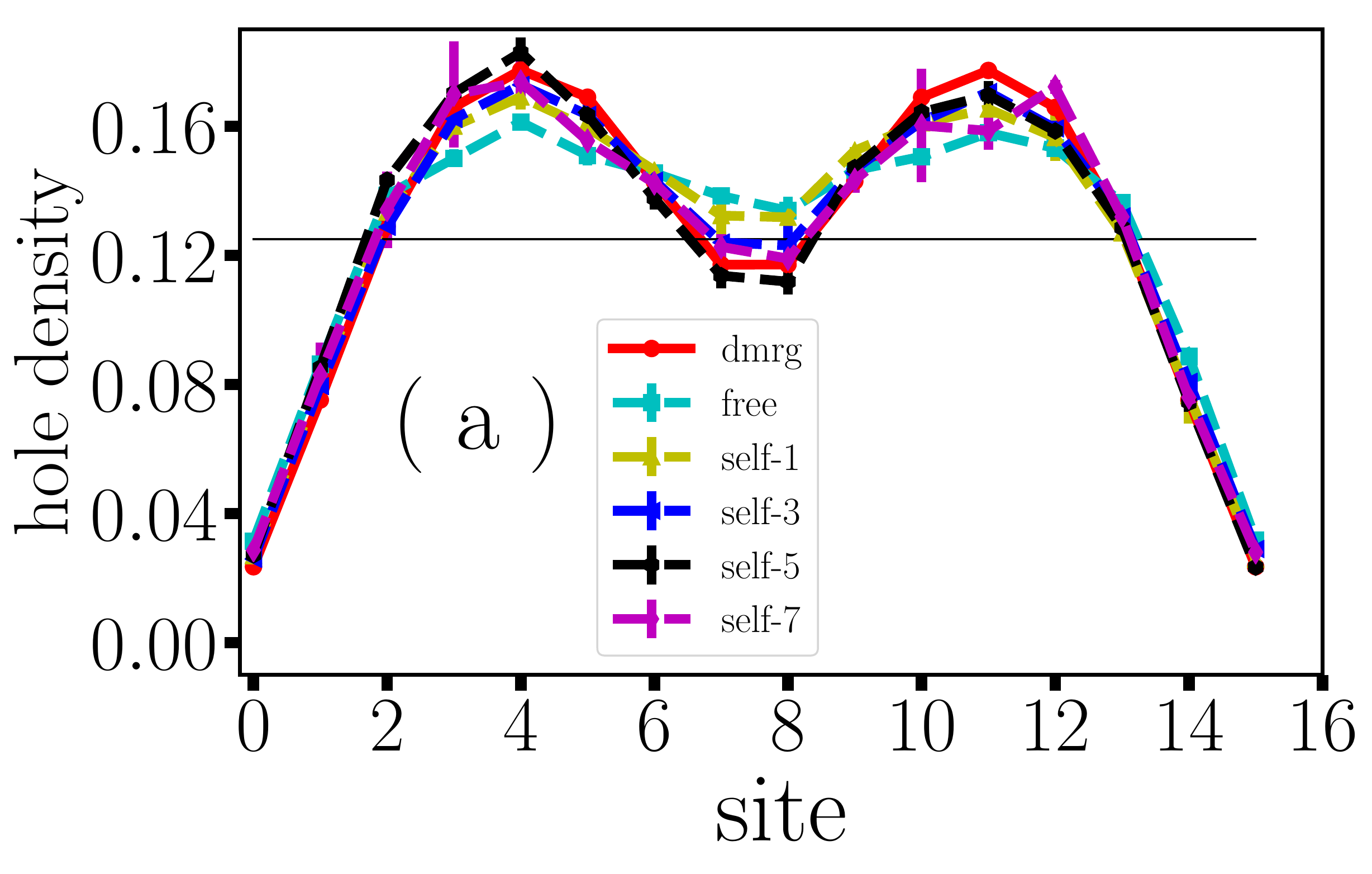}
	\includegraphics[width=82mm]{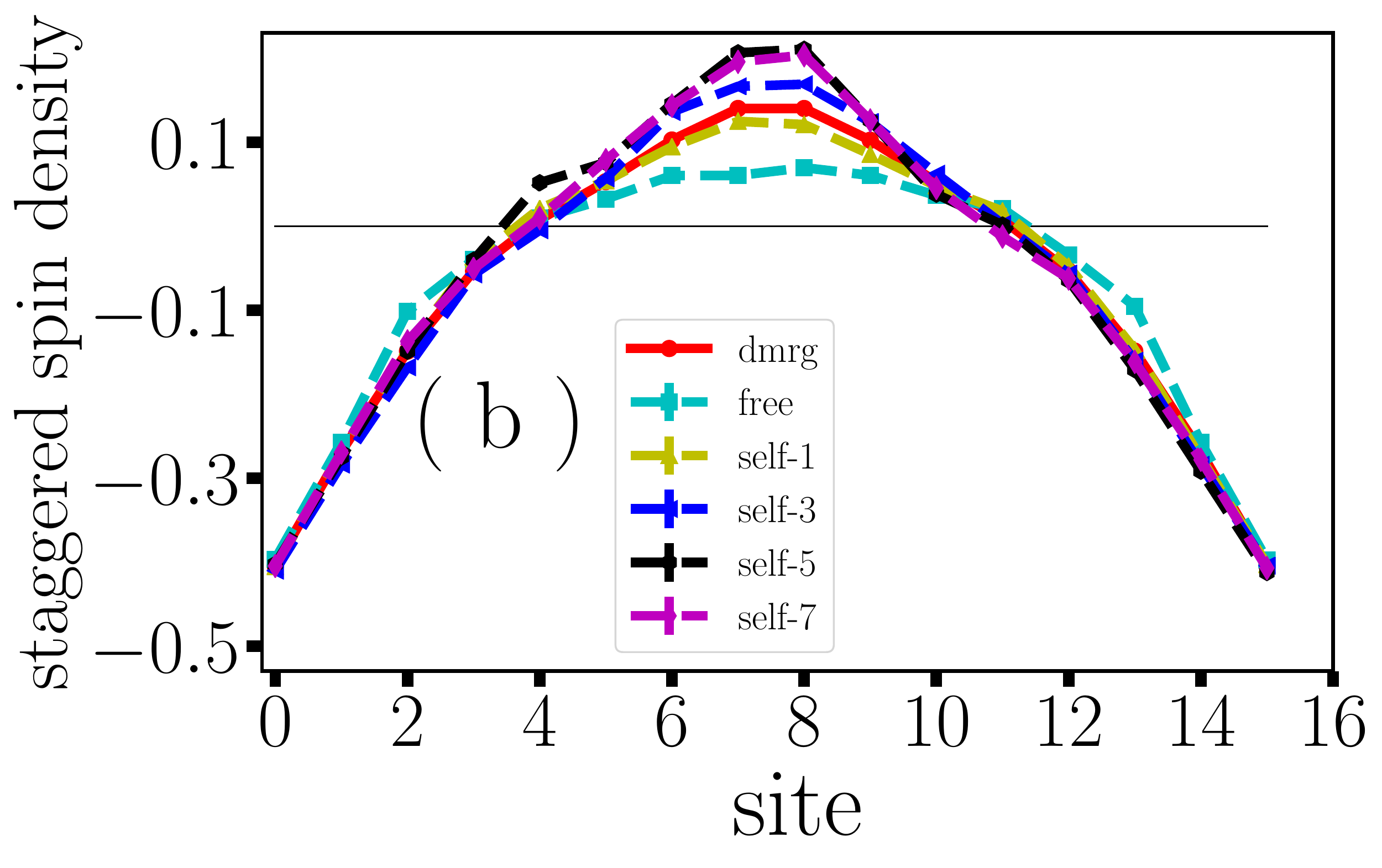}
	\includegraphics[width=80mm]{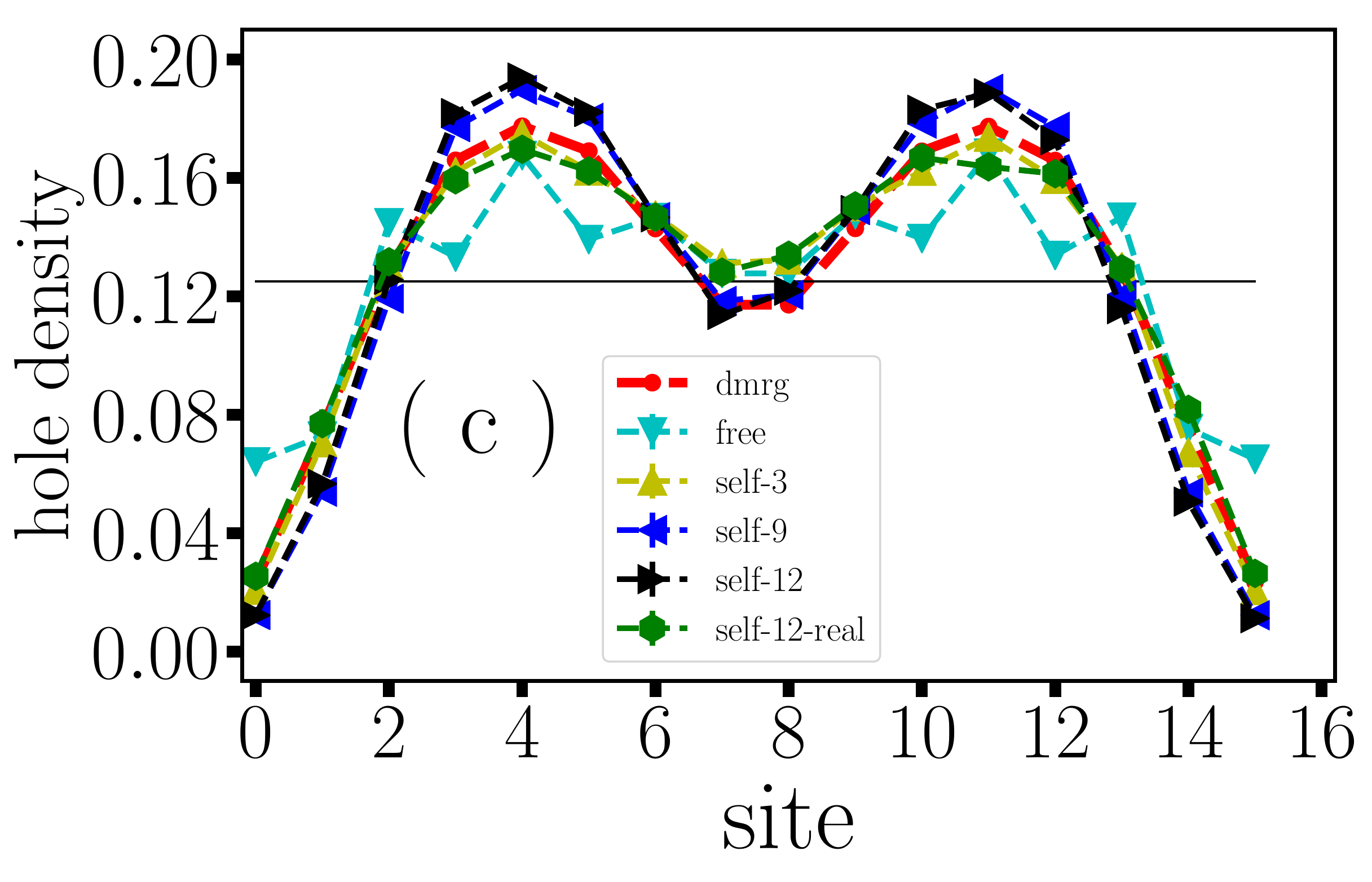}
	\includegraphics[width=82mm]{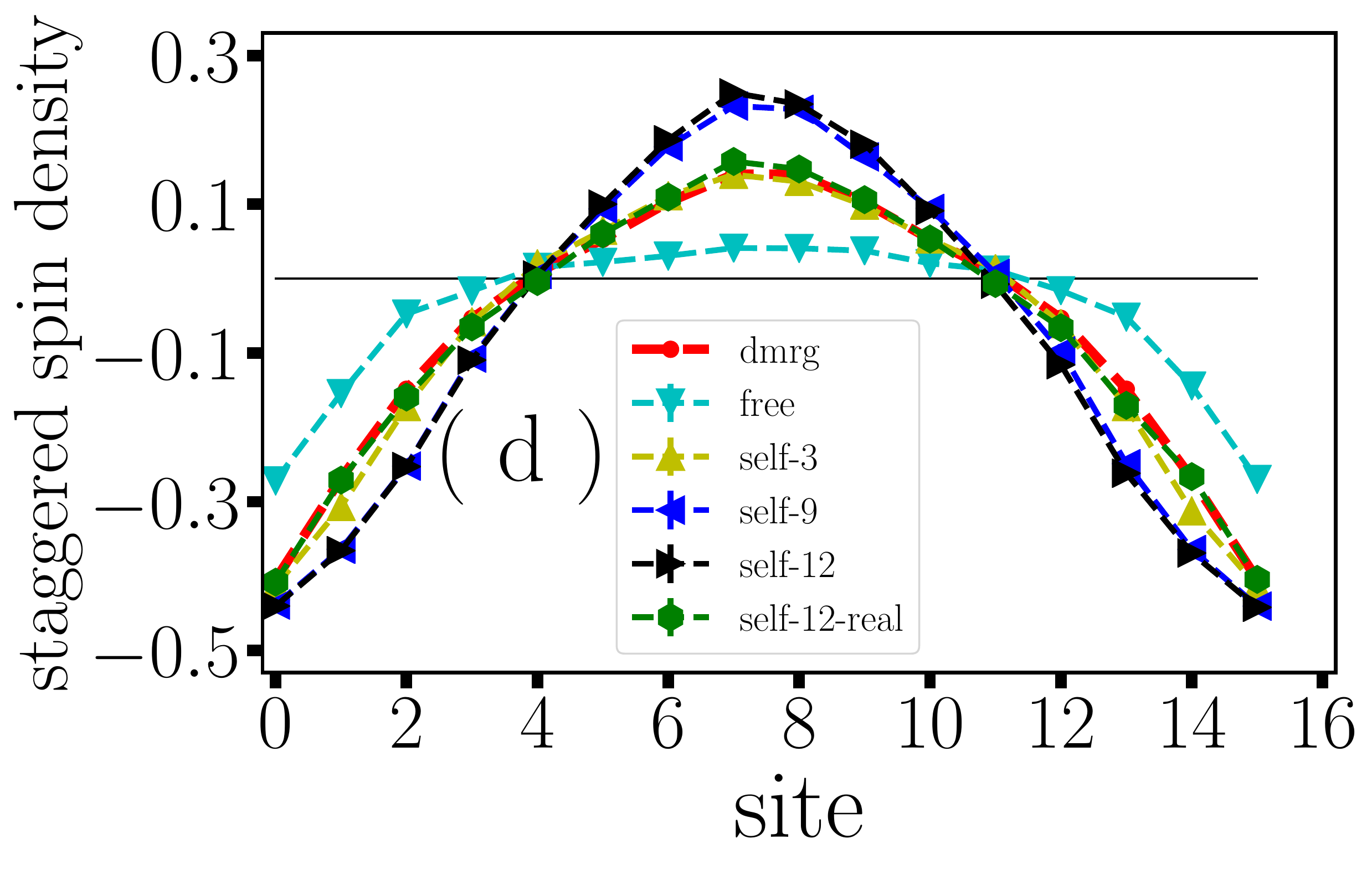}
	\caption{The results of hole and staggered spin density in the self-consistent CP-AFQMC calculation. The numbers in the legend represent the iteration steps in the
		self-consistent process.
		In the upper panel(a) and (b), the real estimator of natural orbital is utilized in the self-consistent calculation.
		In the lower panel (c) and (d), the mixed estimator of natural orbital is used instead. The densities in (c) and (d)
		are also from mixed estimator except the last one marked as ``self-12-real", which are from real estimators in CP-AFQMC using the converged trial wave-function. 
	}
	\label{spin_hole}
\end{figure*}

\section{Model and Method}
\label{model_method}
\subsection{Hubbard Model}
We take the two dimensional doped Hubbard model \cite{Hubbard238,doi:10.1146/annurev-conmatphys-090921-033948,doi:10.1146/annurev-conmatphys-031620-102024} as an example to show the accuracy of the new scheme. The Hamiltonian of Hubbard is  
\begin{equation}
H = K + V = -t \sum\limits_{ \langle i,j \rangle, s} \left(c_{i,s}^\dagger c_{j,s}  + h.c.\right) +U\sum\limits_i n_{i\uparrow}n_{i\downarrow},
\label{eqn:H}
\end{equation}
where $K$ and $V$ represent the kinetic and interacting terms. We focus on the $1/8$ hole doped Hubbard model on a $4 \times 16$ lattice under
cylindrical geometry with $U/t = 8$, for which DMRG \cite{PhysRevLett.69.2863,PhysRevB.48.10345} can provides very accurate results for benchmark. Same as in \cite{PhysRevB.94.235119} we add anti-ferromagnetic pinning fields with strength $h_m = 0.5$ at both edges of the system to calculate the local spin order instead of the more demanding correlation functions.

\subsection{Constrained Path Auxiliary-field Monte Carlo method}
In this subsection, we give a brief introduction of the CP-AFQMC method. More details can be found in \cite{AFQMC-lecture-notes-2013}.
Similar as the power method to calculate the eigenvalue of a matrix with greatest absolute value and the corresponding eigenvector,  
the successively application of the imaginary time evolution operator to an initial state $|\psi_0\rangle$
yields the ground state $|\psi_g\rangle$:
\begin{equation}
|\psi_{g}\rangle \propto \lim_{\beta \rightarrow\infty}e^{-\beta H}|\psi_{0}\rangle
\label{eqn:proj}
\end{equation}
Using trotter-Suzuki decomposition, we can decouple the kinetic and interaction terms in $e^{-\beta H}$ as 
\begin{equation}
e^{-\beta H}=(e^{-\tau H})^{n}=(e^{-\frac{1}{2}\tau K}e^{-\tau V}e^{-\frac{1}{2}\tau K})^{n}+O(\tau^{2})
\end{equation}
with $\beta = \tau n$. The initial state $|\psi_0\rangle$ is usually chosen as a single Slater determinant. The one-body projection term $e^{-\frac{1}{2}\tau K}$ transforms a Slater
determinant to another one, so it is easy to handle. The two-body projection can't be dealt directly. But by taking advantage of the so-called Hubbard-Stratonovich (HS) transformation,
it can be written as an integral or sum of one-body terms over the auxiliary fields. In CP-AFQMC, the wave function is represented as a linear combination of Slater determinants (walkers). 
Eq.~(\ref{eqn:proj}) is then represented as random walks in the Slater determinant space by 
sampling the auxiliary fields.

To attack the negative sign problem in the random walk process, a trial wave-function $|\psi_{T}\rangle$ is introduced in CP-AFQMC. The walkers with negative overlap with
$|\psi_{T}\rangle$ is not allowed to further evolve. Physical quantities commuting with Hamiltonian (e.g., the ground state energy) is calculated as the mixed estimator
\begin{equation}
\langle O\rangle_{\rm mixed}=\frac{\sum_{k}w_{k}\langle\psi_{T}|O|\psi_{k}\rangle}{\sum_{k}w_{k}\langle\psi_{T}|\psi_{k}\rangle}\,,
\label{mix_estimator}
\end{equation}
where $|\psi_{k}\rangle$ is the $k$th walker, $w_k$ is the corresponding weight in the wave-function, and $|\psi_{T}\rangle$ is the trial wave-function. 
For quantities which do not commute with the Hamiltonian, the mixed estimator is biased and
back propagation is applied to calculate the real estimator instead \cite{Zhang97,PhysRevE.70.056702}.

\begin{figure}[t]
	\includegraphics[width=80mm]{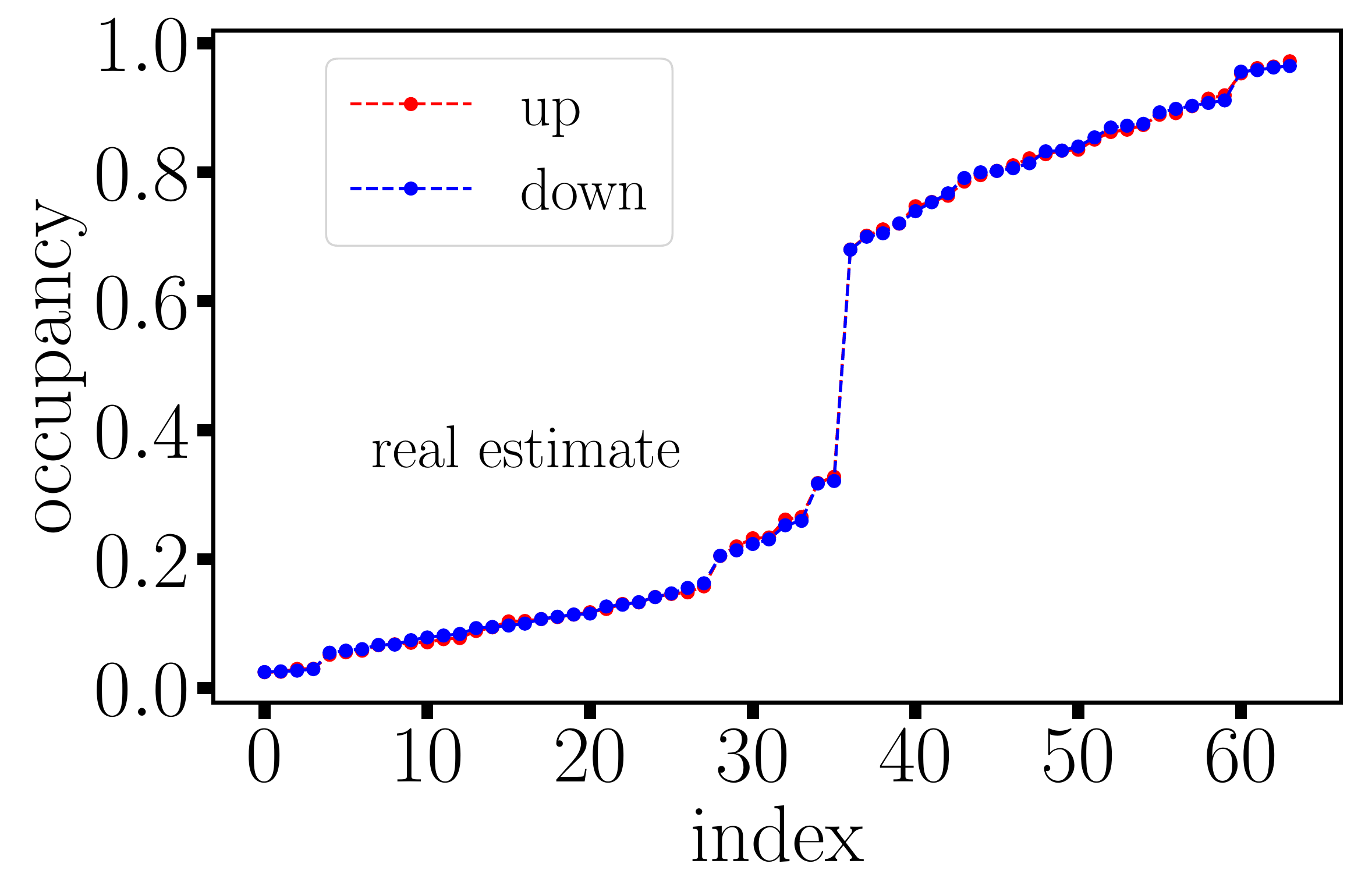}
	\includegraphics[width=80mm]{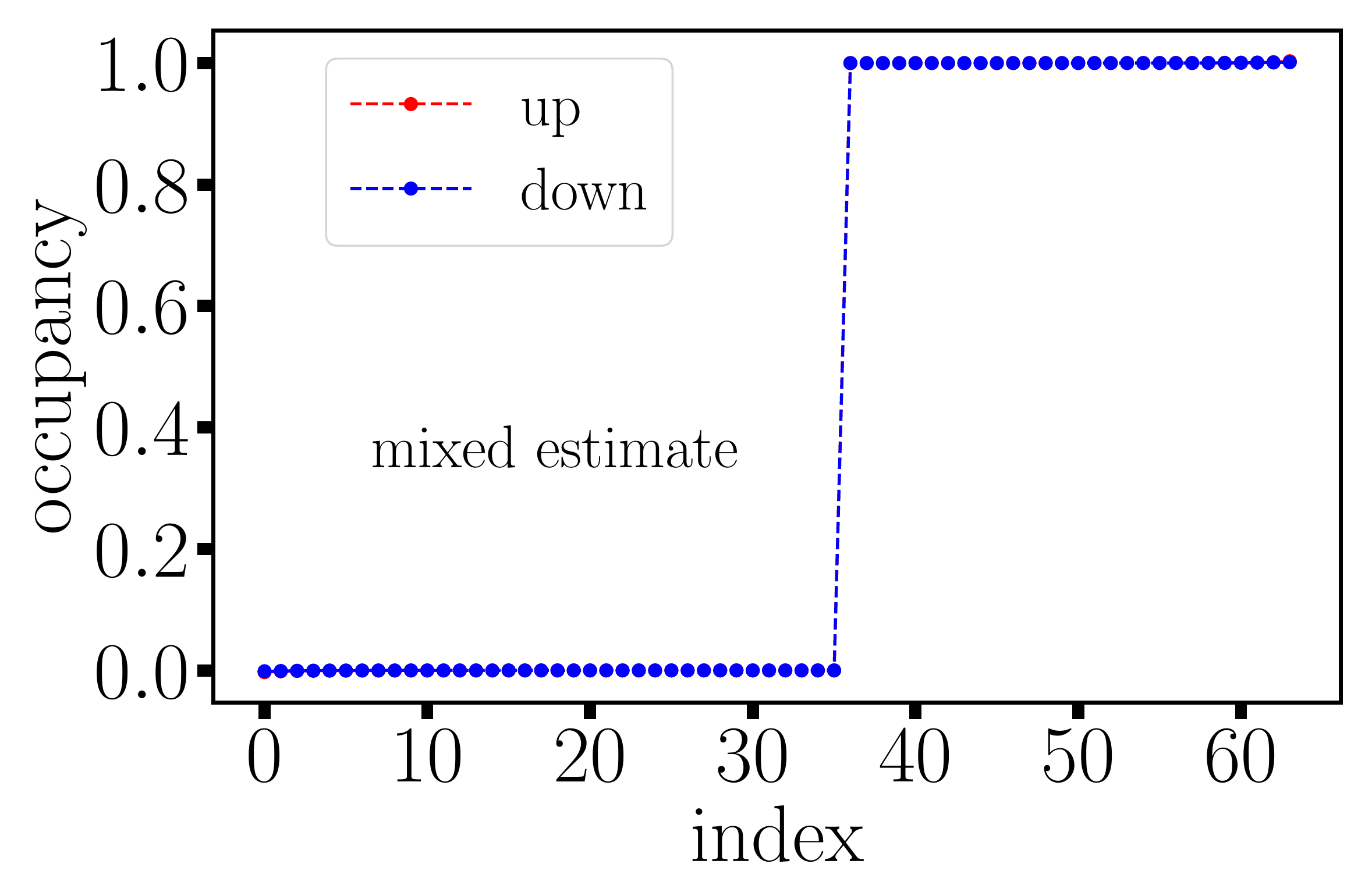}
	\caption{Occupancy for real (upper) and mixed (lower) estimators of the 1-RDM. Only the converged self-consistent results are shown. Notice that the occupancy is either $1$
		or $0$ for mixed estimator.
	}
	\label{occ}
\end{figure}

\section{Self-consistent optimization of the trial wave-function}
As discussed in the above section, the trial wave-function plays a key role in CP-AFQMC and determines the accuracy of the final result.
Trial wave-function were usually chosen empirically \cite{PhysRevB.78.165101,PhysRevB.88.125132,PhysRevB.89.125129,PhysRevB.94.085103} and previous benchmarks on different systems show the constraint errors are modest in many cases.
The idea to optimize the trial wave-function self-consistently was first proposed in 2016 \cite{PhysRevB.94.235119}, followed by generalizations to finite temperature calculation \cite{PhysRevB.99.045108}. In \cite{PhysRevResearch.3.013065},
another self-consistent scheme was developed for the pseudo-BCS type trial wave-function \cite{PhysRevResearch.3.013065}.

Apart from optimizing the trial wave-function self-consistently by coupling the CP-AFQMC calculation with a mean-filed Hamiltonian, it was also proposed to construct new trial wave-function with natural orbitals from CP-AFQMC in the self-consistent process. However, the fluctuation in the natural orbital scheme is very large for the charge density.  
In this work, we propose a new scheme to implement the self-consistent optimization of the trial wave-function in the framework of natural orbitals in CP-AFQMC by taking advantage of the mixed estimators. In this framework,
we first carry out one step of CP-AFQMC calculation with an initial trial wave-function (can be chosen arbitrarily). {In the CP-AFQMC calculation, we can obtain the 1-RDW with the definition below:}
\begin{equation}
{
\rho_{ij}=G_{ij}=\frac{\langle\psi_{g}|c_{i}^{\dagger}c_{j}|\psi_{g}\rangle}{\langle\psi_{g}|\psi_{g}\rangle}
}
\label{def-1rdm}
\end{equation}
{Then by diagonalizing the 1-RDM from QMC calculation,}
\begin{equation}
{
\rho = UVU^{\dagger}
\label{dig_1rdm}
}
\end{equation}
{we can obtain the natural orbitals as the columns of $U$ matrix in Eq.~(\ref{dig_1rdm}).
A new trial wave-function in the form of single Slater determinant (with fixed particle number) can be constructed from the natural orbitals with large occupancy, i.e., from the first $N_e$ (number of electrons) columns of $U$ if the diagonal matrix $V$ in Eq.~(\ref{dig_1rdm}) is in descending order. When spin up and down electrons are decoupled, we need to carry out the above process for both spins.} We then continue the CP-AFQMC calculation with this new trial wave-function and this
procedure is repeated after convergence.

{We compare two schemes to calculate the 1-RDM in CP-AFQMC, i.e., the mixed (see Eq.~(\ref{mix_estimator})) and real estimator from back-propagation \cite{PhysRevE.70.056702}.} The mixed estimator is biased because the 1-RDM don't commute with the Hamiltonian, but it is cheap computationally. The real estimator gives more accurate result for 1-RDM, but it is more demanding computationally \cite{PhysRevE.70.056702}.

\section{Results}
\label{result}

We take the $1/8$ doped Hubbard model with $U = 8$ on a $4 \times 16$ cylinder as an example to test the accuracy of the self-consistent scheme. Same as in \cite{PhysRevLett.99.127004}, anti-ferromagnetic pinning
fields are applied at both edges of the system to enable us to probe the spin correlation by calculating the local spin density.

\subsection{Hole and spin density}

In Fig.~\ref{spin_hole}, we plot the hole and staggered spin density in the self-consistent CP-AFQMC calculation. Only results for one row is shown and the
results for other rows are the same because of the transnational symmetry in the column direction. The
converged (with bond dimension) DMRG results (red) are used as reference. The modulation of spin and hole density agree with the characteristic of the stripe phase \cite{nature_375_15_1995,Zheng1155}. In the upper panel (a) and (b) of Fig.~\ref{spin_hole},
real estimator of natural orbital in CP-AFQMC is utilized in the self-consistent calculation. In the lower panel (c) and (d), the mixed estimator
of natural orbital in CP-AFQMC is used instead. The hole and staggered spin densities in (c) and (d) are also from mixed estimator except the last one marked as
``self-12-real", which are real estimators from CP-AFQMC using the converged trial wave-function.

In Fig.~\ref{spin_hole} (a), we can find very large fluctuations in the hole density even after convergence in the real estimator scheme. But with mixed estimator,
we can find the fluctuation is much smaller and after convergence the real estimator of hole densities using the last step trial wave-function agree well with
the DMRG results in Fig.~\ref{spin_hole} (c). By comparing the staggered spin density in (b) and (d) of Fig.~\ref{spin_hole}, we find the scheme with mixed estimator gives more accurate result
than that with real estimator.

In the self-consistent approach \cite{PhysRevB.94.235119} in which a mean-field Hamiltonian is coupled with the QMC calculation, while very accurate result for spin density was obtained, there exists residual error for the hole density. But in the new scheme with mixed estimator, both spin and hole density agree well with the exact results. 
In the natural orbital scheme, we construct the new trial wave-function from natural orbitals. There is no free parameter in the calculation which makes the results more reliable. Moreover, in the new scheme, the whole
1-RDM was utilized, while in \cite{PhysRevB.94.235119}, only the diagonal elements of 1-RDM is feed back to the construction of new trial wave-function. In the
previous scheme \cite{PhysRevResearch.3.013065} where trial wave-function is constructed also from 1-RDM, real estimator from back propagation \cite{Zhang97,PhysRevE.70.056702} is used which is computational more demanding.

\subsection{Occupancy}

In Fig.~\ref{occ}, we compare the occupancy of converged natural orbitals for the two self-consistent schemes. As shown in Fig.~\ref{occ}, there exist jumps for the occupancy at the position of electron number in the system (counting by the occupancy from large to small) for both schemes. However, for mixed estimator, the occupancies are all 1(0) before (after) the jump (see the appendix for a rigorous proof), while they vary continuously before and after the jump in the real estimator scheme. This result means truncation is needed in the real estimator scheme when constructing a new Slater determinant form 1-RDM ({the 1-RDM from the constructed new trial-wave-function is different from the original 1-RDM from CP-AFQMC calculation}), while all information of 1-RDM is contained in the new trial wave-function in the mixed estimator scheme ({the 1-RDM from the constructed new trial-wave-function is exactly the same as the 1-RDM from CP-AFQMC calculation}). This explains why the hole and spin density from mixed estimator scheme in Fig.~\ref{spin_hole} are more accurate and with less fluctuation.

In the self-consistent scheme in \cite{PhysRevResearch.3.013065}, a pseudo-BCS trial wave-function is also constructed from the QMC result of 1-RDM. It was shown the pseudo-BCS wave-function can also capture the whole occupancy well. But the advantage of the scheme in this work is that we need only to calculate the mixed estimator of the 1-RDM which is much cheaper computationally.

\begin{figure}[t]
	\includegraphics[width=80mm]{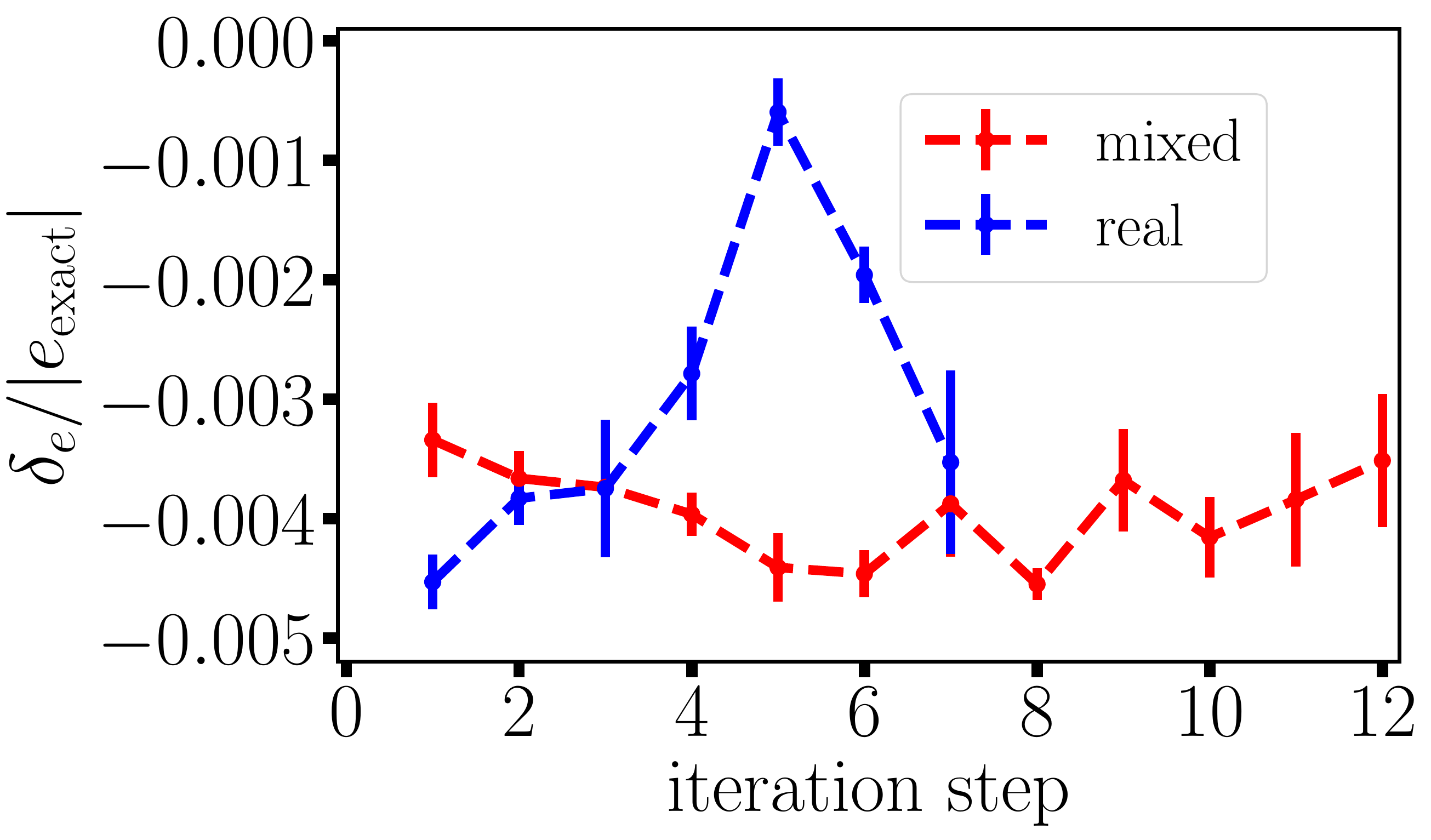}
	\caption{The energies at each iteration step in the self-consistent process for two different schemes.
		The red (blue) means the 1RDM used in the self-consistent process is calculated with mixed (true) estimator. All the energies are calculated
		with mixed estimators to reduce the fluctuation. 
	}
	\label{energy}
\end{figure}

\subsection{Energy}

In Fig.~\ref{energy}, we show the relative error of the QMC energy in each self-consistent step for the two schemes. All the energies in Fig.~\ref{energy} are from mixed estimator to reduce the fluctuation. We can find that the relative error
of ground state energy is about $-0.4 \%$ \cite{foot1} for both the two schemes, which is comparable to the previous self-consistent result \cite{PhysRevB.94.235119}. But the energy in the mixed estimator is with less fluctuation, agreeing with the results of densities in Fig.~\ref{spin_hole}. This result demonstrates the superiority of the mixed estimator scheme again.

\section{summary and perspective}
\label{sum}
In summary, we propose a new scheme to implement the self-consistent optimization of trial wave-functions in CP-AFQMC in the framework of natural orbitals \cite{PhysRevB.94.235119} by taking advantage of the mixed estimators of 1-RDM. Taking the doped
Hubbard model as an example, We compare two schemes in which mixed and real estimator of 1-RDM are calculated in QMC.
We find the scheme with mixed estimator is more accurate, more stable, and more efficient. This new scheme
provides more accurate hole densities than \cite{PhysRevB.94.235119} in which a mean field Hamiltonian is coupled with QMC calculation. It is also more efficient computationally than the pseudo-BCS scheme in \cite{PhysRevResearch.3.013065} because only mixed estimator is needed. This new scheme provides a useful tool for the study of Fermion systems. It is also applicable to the case where up and down orbitals are mixed with spin orbital coupling in the Hamiltonian \cite{PhysRevX.10.031016, xu2023coexistence}.

In the framework of the self-consistent optimization of trial wave-functions by coupling the CP-AFQMC to a mean-field Hamiltonian \cite{PhysRevB.94.235119}, we can
also feed back the cheap mixed estimators to the mean-field Hamiltonian to improve the computational efficiency. But we need to emphasize that in the natural orbital framework with mixed estimators, all information
from 1-RDM is contained in the new wave-function, but in the mean-field scheme, only the diagonal elements of 1-RDM (the densities) are feed to the mean-field Hamiltonian to construct new trial wave-functions.

The scheme in this work can be viewed as a practical realization of the one-body reduced density matrix functional theorem \cite{PhysRevB.12.2111} with high accuracy, which states that the ground state is a functional of the one-body reduced density matrix.  
To further increase the accuracy of CP-AFQMC in the future, we will consider trial wave-function beyond
single Slater determinant \cite{doi:10.1063/5.0031024}, which requires information from two-body reduced density matrix. How to construct a wave-function
consisting of multiple Slater determinants from the cheap mixed estimator will be the key. 

\begin{acknowledgments}
We thank Shiwei Zhang and Hao Shi for useful discussions in related works, and for useful suggestions for this work.
We acknowledges the support
from the National Key Research and Development Program of MOST of China
(2022YFA1405400), the National Natural Science Foundation of China (Grant
No. 12274290) and the sponsorship from Yangyang Development Fund.

\end{acknowledgments}

\bibliography{self_natrual_orbitals.bib}

\begin{thebibliography}{33}
\bibitem[{foo()}]{foot1}
\bibinfo{note}{The energy in CP-AFQMC is not variational.}
\end{thebibliography}

\appendix
\section{The occupancy of the mixed estimated one-body reduced density matrix}
The mixed estimator of the one body reduced density matrix is
\begin{equation}
\rho_{ij}=G_{ij}=\frac{\langle\psi_{T}|c_{i}^{\dagger}c_{j}|\psi\rangle}{\langle\psi_{T}|\psi\rangle}=\frac{\sum_{k}\omega_{k}\frac{\langle\psi_{T}|c_{i}^{\dagger}c_{j}|\phi_{k}\rangle}{\langle\psi_{T}|\phi_{k}\rangle}}{\sum_{k}\omega_{k}}
\label{1brd}
\end{equation}
where $|\psi_{T}\rangle$ is the trial wave-function and $|\psi\rangle$ is the ground state wave-function from QMC calculation.
It is represented as $|\psi\rangle=\sum_{k}\omega_{k}\frac{|\phi_{k}\rangle}{\langle\psi_{T}|\phi_{k}\rangle}$ with $|\phi_{k}\rangle$ the walker in QMC and $\omega_k$ the corresponding weight.
In CP-AFQMC calculation, $|\psi_{T}\rangle$ and $|\phi_{k}\rangle$ are usually chosen as single Slater determinant, so we have
\begin{equation}
\frac{\langle\psi_{T}|c_{i}^{\dagger}c_{j}|\phi_{k}\rangle}{\langle\psi_{T}|\phi_{k}\rangle}=[\phi_{k}(\psi_{T}^{\dagger}\phi_{k})^{-1}\psi_{T}^{\dagger}]_{ji}
\end{equation}
From Eq.~(\ref{1brd}) we have (suppose the weight is normalized, i.e., $\sum_{k}\omega_{k}=1$)
\begin{equation}
\rho_{ij}=\sum_{k}\omega_{k}[\phi_{k}(\psi_{T}^{\dagger}\phi_{k})^{-1}\psi_{T}^{\dagger}]_{ji}
\end{equation}
Let's then calculate $\rho^{2}$ as
\begin{eqnarray}
	\rho^{2} & = & \sum_{k}\omega_{k}[\phi_{k}(\psi_{T}^{\dagger}\phi_{k})^{-1}\psi_{T}^{\dagger}]\sum_{k^{\prime}}\omega_{k^{\prime}}[\phi_{k^{\prime}}(\psi_{T}^{\dagger}\phi_{k^{\prime}})^{-1}\psi_{T}^{\dagger}] \nonumber\\
	& = & \sum_{kk^{\prime}}\omega_{k}\omega_{k^{\prime}}\phi_{k}(\psi_{T}^{\dagger}\phi_{k})^{-1}\psi_{T}^{\dagger}\phi_{k^{\prime}}(\psi_{T}^{\dagger}\phi_{k^{\prime}})^{-1}\psi_{T}^{\dagger} \nonumber \\
	& = & \sum_{kk^{\prime}}\omega_{k}\omega_{k^{\prime}}\phi_{k}(\psi_{T}^{\dagger}\phi_{k})^{-1}\psi_{T}^{\dagger} \nonumber \\
	& = & \sum_{k}\omega_{k}(\sum_{k^{\prime}}\omega_{k^{\prime}})\phi_{k}(\psi_{T}^{\dagger}\phi_{k})^{-1}\psi_{T}^{\dagger} \nonumber \\
	& = & \sum_{k}\omega_{k}\phi_{k}(\psi_{T}^{\dagger}\phi_{k})^{-1}\psi_{T}^{\dagger} \nonumber \\
	& = & \rho
	\label{rho_2}
\end{eqnarray}
Eq.~(\ref{rho_2}) means the eigenvalue of the mixed estimator of one-body reduced density
matrix is either $1$ or $0$. Based on this we can construct a single Slater determinant $\phi$ (normalized) from the natural orbitals of
$\rho$ satisfying:
\begin{equation}
\phi\phi^{\dagger}=\rho
\end{equation}
This means all information of $\rho$ is contained in $\phi$ and 
we can use $\phi$ as the new trial wave-function in the next step CP-AFQMC calculation.
The conclusion is also true if up and down orbitals are mixed with spin orbital coupling in the Hamiltonian \cite{PhysRevX.10.031016,xu2023coexistence}.

\end{document}